\documentclass[aps,prl,twocolumn,superscriptaddress,showpacs,amsmath,floatfix,citeautoscript]{revtex4-2}
\usepackage{amssymb}
\usepackage{amsmath}
\usepackage{colordvi}
\usepackage{amsthm}
\usepackage{subeqnarray}
\usepackage{cases}
\usepackage{enumerate}
\usepackage{graphicx}
\usepackage{epsfig}
\usepackage{epstopdf}
\usepackage{subfigure}
\usepackage{color}
\usepackage{ulem}

\def\avg(#1){\langle#1\rangle}

\def\be{\begin{equation}}
	\def\ee{\end{equation}}
\def\bea{\begin{eqnarray}}
	\def\eea{\end{eqnarray}}

\begin{document}

\title{Interplay between the magnetic structures and the surface states in MnBi$_{2}$Te$_{4}$ from first-principles studies}
\author{Zujian Dai}
\affiliation{CAS Key Laboratory of Quantum Information, University of Science and Technology of China, Hefei 230026, Anhui, China}
\affiliation{Synergetic Innovation Center of Quantum Information and Quantum Physics, University of Science and Technology of China, Hefei, 230026, China}
\author{Gan Jin}
\affiliation{CAS Key Laboratory of Quantum Information, University of Science and Technology of China, Hefei 230026, Anhui, China}
\affiliation{Synergetic Innovation Center of Quantum Information and Quantum Physics, University of Science and Technology of China, Hefei, 230026, China}
\author{Lixin He}%
\email{helx@ustc.edu.cn}
\affiliation{CAS Key Laboratory of Quantum Information, University of Science and Technology of China, Hefei 230026, Anhui, China}
\affiliation{Synergetic Innovation Center of Quantum Information and Quantum Physics, University of Science and Technology of China, Hefei, 230026, China}

\begin{abstract}

The antiferromagnetic (AFM) topological insulator MnBi$_{2}$Te$_{4}$  was believed to have a topological surface state (TSS) with large band gap
due to the ferromagnetic (FM) order on the surface, and be able to host the long-sought axion states.
However, recent angle-resolved photoemission spectroscopy (ARPES) experiments indicate that the TSS is gapless,
contradicting the theoretical predictions. Meanwhile, several experiments have suggested that there is robust out-of-plane FM order on the surface of MnBi$_{2}$Te$_{4}$. To understand these seemingly contradictory results, we carry out comprehensive first-principles calculations to investigate the interplay between the surface magnetism and the TSS. Our calculations provide direct evidence that in a wide range of parameters,
the (nearly) gapless TSS can coexist with the surface FM order, therefore solving the paradox of the surface magnetism and the gapless TSS. We further show that proximity effects can be a promising route to open the gap in the TSS of MnBi$_{2}$Te$_{4}$. Our research deepens the understanding of the relationship between surface magnetism and TSS.

\end{abstract}
\maketitle

%%%%%%%%%%%%%%%%%%%%%%%%%%%%%%%%%%%%%%%% INTRODUCTION  %%%%%%%%%%%%%%%%%%%%%%%%%%%%%%%%%%%%%%%%%%%%%%%%%%%%%%%%%%%

%\section{introduction}

  The interplay between magnetism and nontrivial band topology may lead to rich physical
   phenomena that not only are interesting for fundamental physics
  but also have important potential applications in spintronic devices  \cite{smejkal_topological_2018,tokura_magnetic_2019}.
  Therefore, MnBi$_{2}$Te$_{4}$, the first synthesized
  intrinsic magnetic topological insulator (MTI)\cite{Jole&Moor_MTI_2010,Bernevig_MTI_2013,Leon_MTI_2011}, has attracted great attention since its appearance~\cite{Otrokov_MnBiTe_2019_1,Otrokov_MnBiTe_2019_2,Xuyong_2019,Wangjing_2019,Liuqihang_2019}.
 MnBi$_{2}$Te$_{4}$ has an A-type antiferromagnetic (AFM) structure,
which enables unique thickness-dependent topological properties:
 for a thin film with an odd number of layers, it is the QAH insulator
 \cite{wangjing_QAHE_2020,zhangyuanbo_2020,liu_axion_2020,liu_axion_2021,deng_QAHE_2021},
whereas for a thin film with an even-number of layers, it is the long-sought axion insulator\cite{Zhangshoucheng_2008,Vanderbilt_axion_2018,chen_axion_2019,nenno_axion_2020,
 sekine_axion_2021,Zhanghaijun_axion_2020,Wangjing_2019,
 Benjamin_and_Vishwanath_axion_2020,liu_axion_2020,gao_layer_2021,liu_axion_2021,Liuqihang_axion_2021}.

However, despite intensive research, the nature of the topological surface states (TSSs) of this material is still very controversial\cite{Otrokov_MnBiTe_2019_1,Reinert_gapped_2019,maozhiwiang_gapped_2019,zeugner_gapped_2019,ChenYJ_gapless_2019,liuqihang_gapless_2019,dinghong_gapless_2019,Adam_gapless_2020,chen_axion_2019,Johnson_gapless_2020,
chenxh_gapless_2020,hejf_gapless_2020,zvezdin_gapless_2021,hou_gapless_2020,Liuqihang_gapless_2021,garnica_gapless_2022}.
First-principles calculations predicted that MnBi$_{2}$Te$_{4}$ has a TSS with considerable energy gap (larger than 60 meV) due to the
ferromagnetic (FM) spin order on the surface\cite{Otrokov_MnBiTe_2019_1,Wangjing_2019,Xuyong_2019,maozhiwiang_gapped_2019,Reinert_gapped_2019,zeugner_gapped_2019},
which is promising for achieving QAH at rather high temperatures. Nevertheless, the zero-field QAHE was observed only at rather low temperatures (below 1.6 K) in this system \cite{liu_axion_2020, zhangyuanbo_2020}.
The gapped TSS was reported in early ARPES experiments \cite{Otrokov_MnBiTe_2019_1,Reinert_gapped_2019,maozhiwiang_gapped_2019,zeugner_gapped_2019}.
However,  more recent ARPES measurements have observed  a nearly perfect Dirac cone or
a strong reduction of the gap at the Dirac point on the MnBi$_{2}$Te$_{4}$ (0001) surface
   \cite{ChenYJ_gapless_2019,liuqihang_gapless_2019,dinghong_gapless_2019,
   Adam_gapless_2020,chen_axion_2019,Johnson_gapless_2020,chenxh_gapless_2020,hejf_gapless_2020,zvezdin_gapless_2021,hou_gapless_2020,garnica_gapless_2022}.

To understand the origin of the gapless TSS~\cite{Liuqihang_gapless_2021}, three scenarios have been proposed, including surface magnetic reconstruction
\cite{chen_axion_2019,Adam_gapless_2020,yuan_gapless_2020}, geometric reconfiguration \cite{hou_gapless_2020,xiexc_gapless_2020,shikin_gapless_2020,du_gapless_2021,Liuqihang_gapless_2021_1} and hybridization of the surface and bulk bands  \cite{chenchaoyu_gapless_2020,yangshoulong_gapless_2021}.
For surface magnetic reconstruction, it has been shown that three types of surface spin reorientation may led to gapless TSSs, including paramagnetism (PM), in-plane A-type AFM, and G-type AFM\cite{liuqihang_gapless_2019}.
However, both time-resolved angle-resolved photoemission spectroscopy (ARPES)\cite{McQueeney_support_2020},
and  magnetic force microscopy \cite{wuweida_support_2020}
suggested that there is robust out-of-plane FM order on the surface of MnBi$_{2}$Te$_{4}$.
A key perplexity here is weather the gapless TSSs can coexist with surface FM order\cite{McQueeney_support_2020}.

 In this Letter, we carry out comprehensive first-principles calculations to investigate the interplay between the surface
 magnetism and the TSS. Our calculations provide solid evidence that in a wide range of parameters, the (nearly) gapless TSSs
can coexist with the surface FM order, therefore solving the paradox of the surface magnetism and the gapless TSS.
We further show that proximity effects can be a promising route to open the gap in the TSSs of MnBi$_{2}$Te$_{4}$.

\begin{figure*}[ht]
	\centering
	\includegraphics[width=0.9\textwidth]{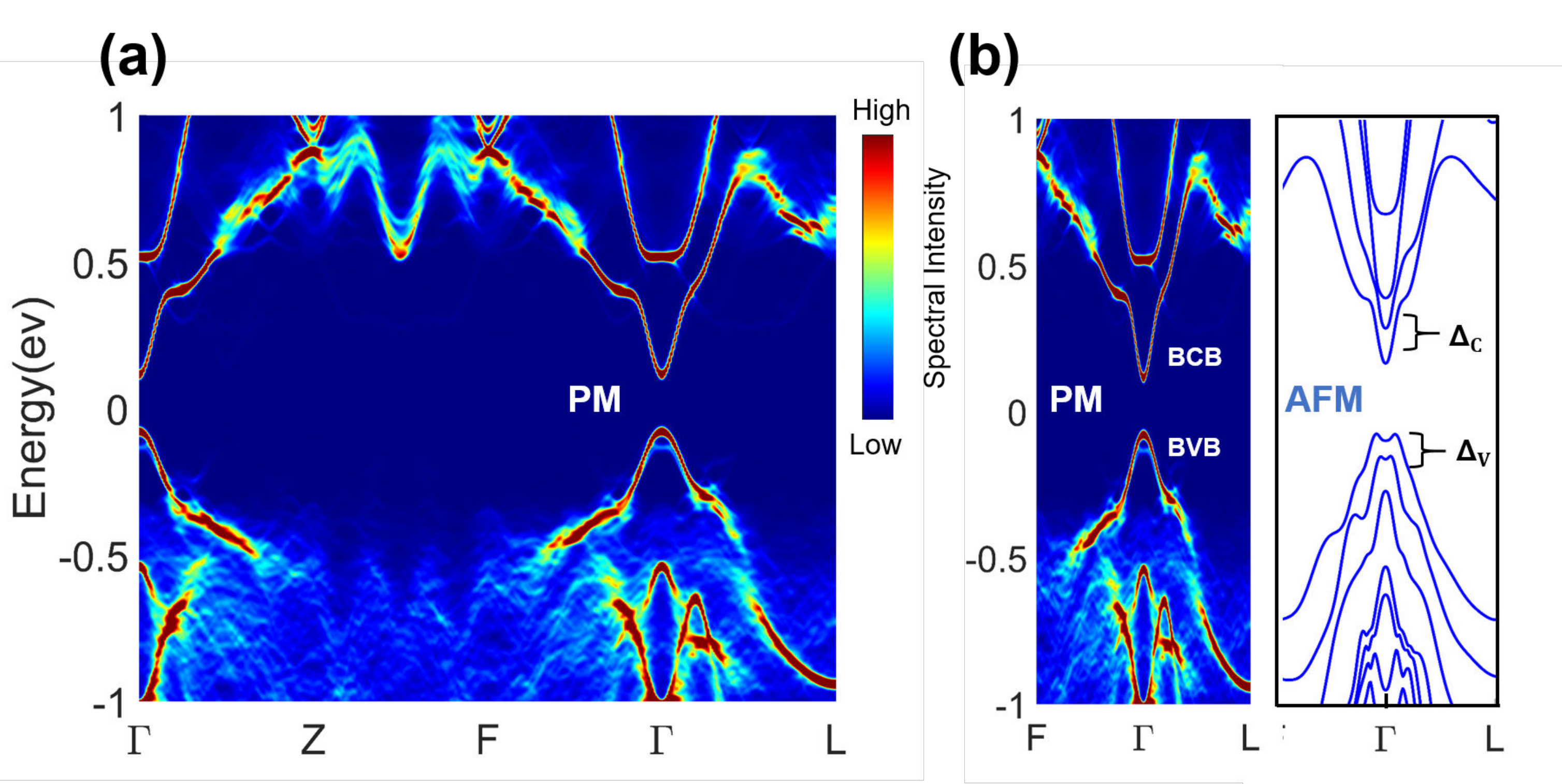}
	\caption{ (a)The unfolded band spectra of MnBi$_{2}$Te$_{4}$ in the PM phase.
		(b) Comparison of unfolded band spectra of MnBi$_{2}$Te$_{4}$ in the PM and A-type AFM phases around the $\Gamma$ point.
		The color shows the spectra intensity, which is broadened by 4 meV.
		BCB and BVB denote the bulk conduction bands, and bulk valence bands respectively.
	    The energy band splitting of BCB and BVB in the AFM phase is denoted by $\Delta_{\text{C}}$ and $\Delta_{\text{V}}$, respectively. }
	\label{fig:paramagnetic bulk}
\end{figure*}

%\section{Computational details}

The first-principle calculations are carried out with the Atomic orbtial Based Ab-initio Computation at UStc (ABACUS) code\cite{Chen_mohan2010,Lipengfei2016}.
The Perdew-Burke-Ernzerhof\cite{Perdew1996} (PBE) exchange-correlation functional is adopted and
the DFT-D3 correction is used to account the van der Waals (vdW) interactions\cite{Grimme2010}.
A Hubbard-like $U$ value of 4.0 eV is used for the half-filled, strongly localized Mn 3d orbitals \cite{Sutton1998,qu_dft_2022}.
The ABACUS code is developed to perform large-scale density functional theory calculations based on numerical atomic orbitals (NAO)\cite{Chen_mohan2010}.
The optimized norm-conserving Vanderbilt (ONCV) \cite{Hamann2013}
fully relativistic  pseudopotentials \cite{Theurich2001} from the PseudoDojo library\cite{Van2018} are used.
The valence electrons for Mn, Bi and Te are
3s$^{2}$3p$^{6}$3d$^{5}$4s$^{2}$, 5d$^{10}$6s$^{2}$6p$^{3}$ and 4d$^{10}$5s$^{2}$5p$^{4}$, respectively, and
the NAO bases for Mn, Bi and Te are 4s2p2d1f, 2s2p2d and 2s2p2d, respectively.
In the self-consistent and band structure calculations, the energy cutoff for the wave functions is set to 120 Ry.
Experimental lattice parameters \cite{lee_crystal_2013} have been used. The atomic positions are fully optimized until all forces are less than 0.01 eV/\AA.

%\section{Results and discussion}
%\subsection{Band structures of bulk MnBi$_{2}$Te$_{4}$ in AFM and paramagnetic phases}

The MnBi$_{2}$Te$_{4}$ has a vdW stacking structure (space group $R\bar{3}m$) with a MnTe bilayer sandwiched by
Bi$_{2}$Te$_3$. The unit cell consists of a Te-Bi-Te-Mn-Te-Bi-Te septuple layer (SL)\cite{Ziya_MnBiTe_2019,zeugner_gapped_2019,Yan_MnBiTe_2019}.
Below the N\'{e}el temperature $T_N = 25 \text{K}$, the spins within each SL are found to be parallel to the
out-of-plane easy axis but antiparallel within the adjacent SLs\cite{Otrokov_MnBiTe_2019_1,zeugner_gapped_2019,Yan_MnBiTe_2019,Sales_MnBiTe_2019}.
The conduction and valance band splittings in MnBi$_{2}$Te$_{4}$ below the N\'{e}el temperature were reported, due to the spins ordering
below the N\'{e}el temperature \cite{ChenYJ_gapless_2019,yangshoulong_gapless_2021}.

We first examine the impact of magnetism on the bulk band structures, by comparing the band structures
of MnBi$_{2}$Te$_{4}$ in the PM and A-type AFM phases.
To simulate the band structure of the bulk MnBi$_{2}$Te$_{4}$ in the PM phase, we construct
 a 4$\times$4$\times$4 supercell. The Ising-like spins on Mn ions \cite{McQueeney_support_2020} are randomly initialized.
 We neglect the structure relaxation due to different spin configurations.

The band structures calculated in the supercell are unfolded to the nonmagnetic unit cell\cite{dai_unfolding_2022,SM}.
We find that the unfolded band structures of different random spin configurations are almost the same,
which suggests that the supercell is large enough to describe the PM state.
The unfolded band structures of the PM phase are shown in
Fig.~\ref{fig:paramagnetic bulk}(a), which shows a sharp
and well-defined band spectra around the $\Gamma$ point.
However, the bands become blurred around the $Z$ and $F$ points.
This is because, around the $\Gamma$ point, where the Bloch states have long wave length,
the spin configurations are well averaged within the wave length range. In contrast, when the wave length of the Bloch states is short, the short range
spin fluctuation blurs the band structure.

The band structures of MnBi$_{2}$Te$_{4}$ in the PM phase around the $\Gamma$ point are compared to those in the AFM phase
in Fig.~\ref{fig:paramagnetic bulk}(b).
In addition to some differences in details, there is a significant difference between the AFM bands and PM bands, i.e.,
the highest valence band and the lowest conduction bands of PM states split into two bands  in the AFM states.
The energy splitting of the conduction band minimum(CBM) is approximately 90 meV, which is larger than
that from ARPES experiments\cite{ChenYJ_gapless_2019} (50 meV at 7.5 K).
The difference might come from that the magnetic state near the surface is not perfect AFM in experiments.

%\subsection{Band structure of surface states} 	

We now turn to the surface states of the MnBi$_{2}$Te$_{4}$ under different magnetization.
We construct a slab containing 4$\times$4$\times$6 MnBi$_2$Te$_4$ unit cells.
There are 96 Mn atoms in the slab. We change the spin orientations of Mn ions in the slab
to realize different magnetic configurations.
A 15 \AA ~vacuum is added to avoid the interactions between the slab and its periodic images.
The atomic positions of the slab is relaxed under the A-type AFM configuration.

 \begin{figure}
 	\centering
 	\includegraphics[width=0.5\textwidth]{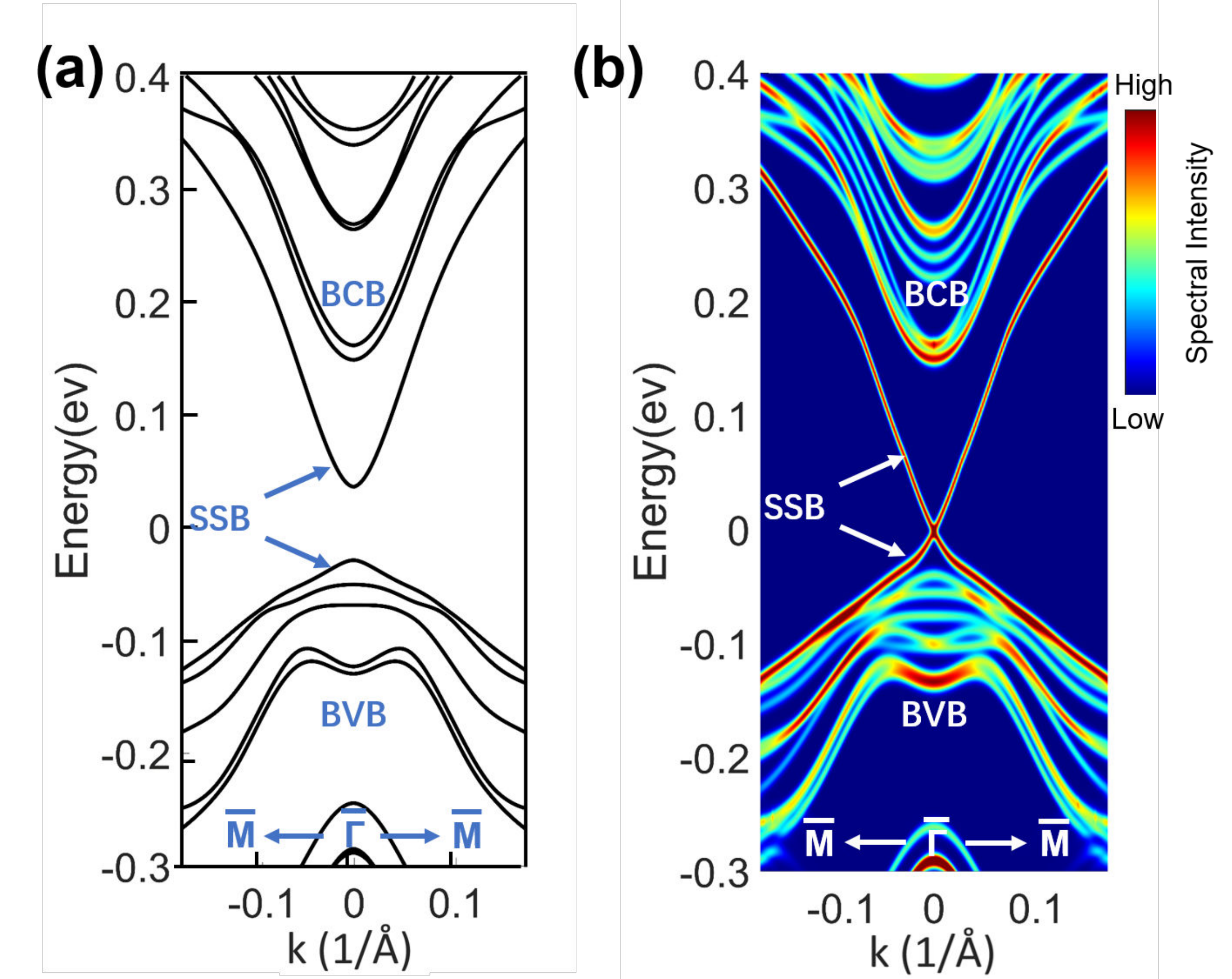}
 	\caption{ (a) The band structure of the  1$\times$1$\times$6 AFM slab.
 		(b) The unfolded band spectral of the 4$\times$4$\times$6 PM slab.
 		The color shows the intensity of the unfolded band spectra, which is broadened by 4 meV.
 	  SSB and BCB and BVB denote the surface-state bands, bulk conduction bands, and bulk valence bands respectively.}
 	\label{fig:paramagnetic slab}
 \end{figure}

The band structure of the perfect  A-type AFM MnBi$_{2}$Te$_{4}$ slab is shown in  Fig.~\ref{fig:paramagnetic slab}(a).
We determine the surface states by analyzing the real space distributions of the  wave functions.
The surface state has a considerable gap of 65 meV,
which is consistent with previous theoretical results
\cite{Otrokov_MnBiTe_2019_1,Wangjing_2019,Xuyong_2019,maozhiwiang_gapped_2019,Reinert_gapped_2019,zeugner_gapped_2019}.

When the temperature is above the N\'{e}el temperature,
the experiment shows that MnBi$_{2}$Te$_{4}$ becomes PM
and shows gapless surface states.
It has been predicted that the PM MnBi$_{2}$Te$_{4}$
is a strong topological insulator with a gapless TSS in the  (0001) direction
\cite{ChenYJ_gapless_2019,Liuqihang_2019}.
We calculate the (unfolded) band structures of the PM slab, which are shown in Fig.~\ref{fig:paramagnetic slab}(b).
As expected, the unfolded spectra of the PM slab show a gapless TSS with a
bright Dirac point.
We find that the surface states are mostly localized on the outermost two layers, which is consistent
with model Hamiltonian calculations\cite{xiexc_gapless_2020}.
Therefore, one may expect that the magnetism of the top two SLs may have a great influence on the TSS.

Previous theoretical investigations suggested that the gapped surface states in the AFM MnBi$_{2}$Te$_{4}$ may protect the axion states\cite{Zhanghaijun_axion_2020,Wangjing_2019,
	Benjamin_and_Vishwanath_axion_2020,liu_axion_2020,gao_layer_2021,liu_axion_2021,Liuqihang_axion_2021}.
However, experimentally, it has been found that the surface states of the MnBi$_{2}$Te$_{4}$ are gapless even below the N\'{e}el temperature,
which contradicts to the theoretical predictions. The origin of the gapless surface states is under intensive discussion\cite{ChenYJ_gapless_2019,liuqihang_gapless_2019,
dinghong_gapless_2019,Adam_gapless_2020,chen_axion_2019,Johnson_gapless_2020,
chenxh_gapless_2020,hejf_gapless_2020,zvezdin_gapless_2021,hou_gapless_2020,Liuqihang_gapless_2021,garnica_gapless_2022}.
It has been suggested that the gapless TSS may come from the magnetic reconstructions at the surface.
To understand the surface states in MnBi$_{2}$Te$_{4}$, we calculate the surface states under different surface magnetizations.

\begin{figure}[t]
	\centering
	\includegraphics[width=0.5\textwidth]{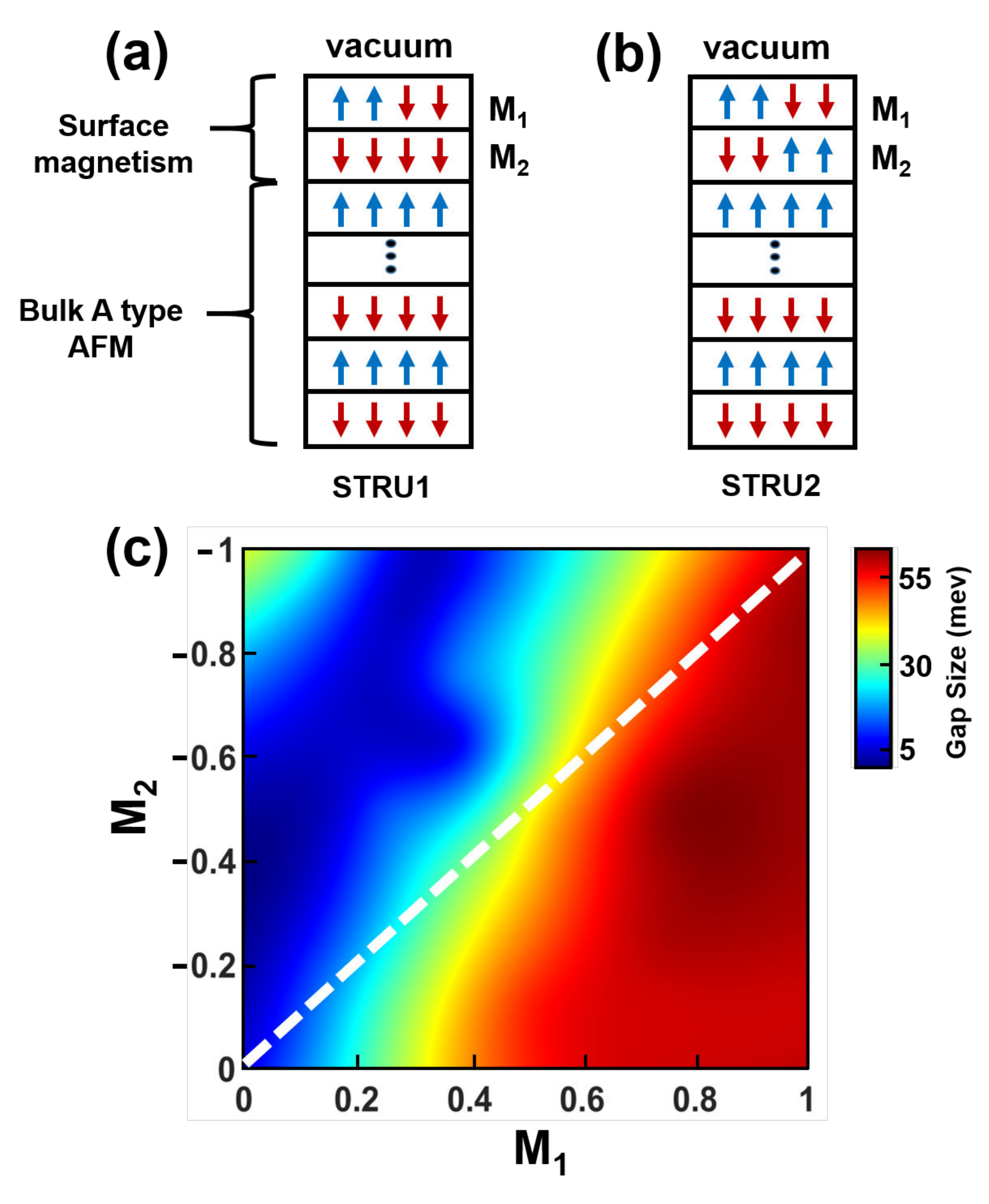}
	\caption{ Illustration of two different of surface magnetic reconstructions.
		(a)  The first SL are set to PM, i.e., $M_1 = 0$ ;
		(b)  Both the first and the second SLs are set to PM, i.e.,  $M_1$=$M_2$= 0.
		(c)  The band gap of the TSS under different surface magnetization $M_{1}$ and $M_{2}$ . The color shows the size of the gap. The white dashed line is the guideline for $M_{1}$=$-M_{2}$.  }
	\label{fig:magnetic configurations}
\end{figure}

We first calculate the surface states on a system, in which all other layers are in the A-AFM state,
except that the top SL is PM as sketched in Fig.~\ref{fig:magnetic configurations}(a).
We focus on the TSS of the top surface, which can be easily identified by its localization.
We find that the TSS of the top surface is gapped by approximately 38 meV, slightly larger than half of the gap of the idea FM surface.
This is because the TSS are most localized on the first two SLs of the surface \cite{SM},
and the FM magnetic order in the second SL still has large effect on the TSS, which opens the gap.
To examine this idea, we calculate the TSS of the system,
where the top two SLs of the slab are set to PM as sketched in Fig.~\ref{fig:magnetic configurations}(b).
Indeed, the band gap of the TSS of the top surface is approximately 0.6 meV,
which is too small to be detected experimentally.
These results suggest that we must consider the magnetic reconstructions of the first two SLs to
understand the TSS.

To obtain a more general understanding of the relationship between the surface magnetism and the TSS,
we systematically calculate the TSS of systems with different magnetizations of the top two SLs.
The surface magnetic reconstruction has several possible microscopic origins.
For example, it may come from the competition of the magnetic interactions of Mn-layers~\cite{McQueeney_support_2020},
or it may come from the native point defects in MnBi$_{2}$Te$_{4}$~\cite{hou_gapless_2020,du_gapless_2021,garnica_gapless_2022}.
It may also come from the domain wall structures of MnBi$_{2}$Te$_{4}$\cite{wuweida_support_2020}.
In this work, we do not discuss the mechanisms of the magnetic reconstructions,
which would be an interesting topic for future investigations.

We define the layer magnetization of each SL as,
\begin{equation}
M_{i}=\frac{N_{i,\uparrow}- N_{i,\downarrow}}{N_{i,\uparrow} +N_{i,\downarrow}},
\end{equation}
where $N_{i,\uparrow}$ ($N_{i,\downarrow}$) is the total number of spins up (down) in the $i$-th SL.
Because the interlayer exchange interactions are AFM-like, we set the layer magnetizations $M_{1}$ and $M_{2}$ to have opposite signs.
The value of the TSS gap as a function of layer magnetization $M_{1}$ and $M_{2}$  is shown in Fig.~\ref{fig:magnetic configurations}(c).
When both $M_{1}$ and $M_{2}$  are zero (i.e., they are both PM), the band gap is approximately zero, whereas when $M_{1}$=-$M_{2}$=1 (i.e., perfect
A-type AFM),
the energy gap of TSS is approximately 65 meV. Remarkably,  when $|M_{2}|$ is larger than  $|M_{1}|$,
the surface states tend to have very small band gaps. Especially, around the $M_{1} \approx  -{1 \over 3} M_{2}$, the
band gap of TSS is nearly zero. In general, there is a large region in which the energy gap size is tiny.
% (\red{$<$ 5 meV?) {\it Is 5 meV small enough?}}.
In this region, the total surface magnetization $M_1$+$M_2<$0, i.e., there is net FM magnetization.
Interestingly, when $M_1$+$M_2$=0, i.e., the white dashed line in Fig.~\ref{fig:magnetic configurations}(c), the band gap of TSS is not necessarily zero, which increases with increasing $M_1$.
When $|M_{1}|>M_{2}|$, the TSS tends to a have large gap. Especially, near $M_1$=0.8, $M_2$=-0.45, the gap is approximately 69 meV, which is even slightly larger than that of the perfect A-type AFM slab.
Because $M_2$ is in the inner layer which is more affected by the bulk magnetism,
it is highly plausible that the system is in the region of $|M_{2}|>|M_{1}|$, where the TSS gap is small.
These results may well explain why MnBi$_{2}$Te$_{4}$ has gapless surface states.
The unfolded spectra of surface states under different surface magnetization can be found in the Supplement Materials \cite{SM}.

According to our calculations, the magnetization of the surface layers does not has to be zero to
have a (nearly) gapless TSS.
The magnetization of the first SL is weakened by dangling bonds or defects\cite{hou_gapless_2020,chenxh_gapless_2020,wuweida_2020,garnica_gapless_2022},
and is much weaker than that of the second SL.
Therefore the first two SLs still have large net FM magnetization, while the gap of the TSS is nearly zero.
This scenario is different from previous works, where the magnetic reconstruction leads to zero magnetization on the surface
\cite{chen_axion_2019,Adam_gapless_2020,yuan_gapless_2020}.
Our results provide direct evidence that the gapless TSS may coexist
with the FM order on the surface\cite{McQueeney_support_2020,wuweida_support_2020}.
These results also suggest that the interaction between the TSS and surface magnetism is too complex to be
described well by effective models, and atomistic models are necessary to obtain the quantitative results.

\begin{figure}[t]
	\centering
	\includegraphics[width=0.5\textwidth]{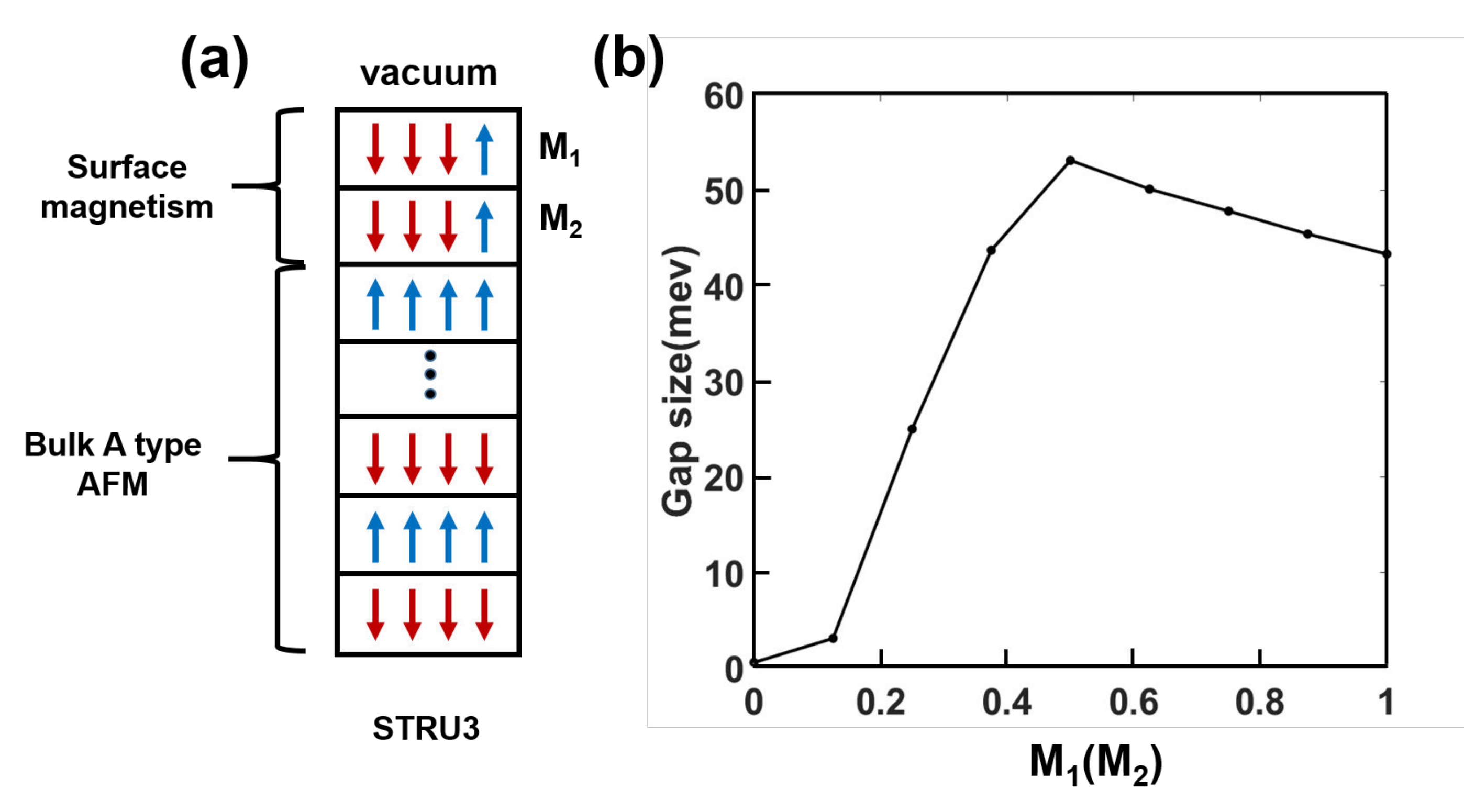}
	\caption{  (a) Illustration of the magnetic configurations with the changes  of
		the magnetization of the first two layers and let $M_1 = M_2 $;
		(b) The TSS band gap as a function of the surface magnetization $M_{1} $=$M_{2}$.}
	\label{fig:open gap}
\end{figure}

%\subsection{How to get a gapped surface states}

To realize the QAH effect and axion insulator, we need to open a considerable gap in the TSS.
A promising way to open the gap is to polarize the spins of outermost layers using the proximity effects.
To study the relation between the spin polarization and TSS band gap in this case,
we calculate the band structures using the spin configurations shown in Fig.~\ref{fig:open gap}(a),
i.e., the first two SLs are polarized
along the same direction.
To simplify the discussion, we assume $M_{1}$=$M_{2}$.
The calculated band gap as a function of $M_1$ is shown in Fig.~\ref{fig:open gap}(b).
The TSS gap first increases with  increasing $M_1$ and saturates at $M_1$=0.5, with a TSS gap of  55 meV.
The gap decreases slightly when further increasing $M_1$.
The decrease in the TSS gap may come from the tendency to the Weyl semimetal\cite{Xuyong_2019,Wangjing_2019,xuyong&&duanwenhui_2019} when all spins
are polarized in the same direction, i.e., a FM state.
Inn practice, $M_1$ could be larger than $M_2$, or even has opposite sign of $M_2$
because the first SL is closer to the magnetic materials,
but according to Fig.~\ref{fig:magnetic configurations}(c), we could still expect that the TSS may have a considerable band gap.
Therefore, it is promising to open a robust gap in MnBi$_2$Te$_3$,
through the heterojunction technique and the proximity effect \cite{wijiazhen_hetro_2019,hou_axion_2019,Yanbinghai2020,Wu2020,Canali_hetro_2021}.
	
%\section{Summary}

To summarize, we carry out a comprehensive investigation of the interplay between the surface states and magnetism of MnBi$_{2}$Te$_{4}$.
Our calculations provide direct evidence that in a wide range of parameters, the gapless (or heavily reduced gap) TSS can coexist with the surface FM order.
We further show that  proximity effects can be a promising route to open the gap of the TSS of MnBi$_{2}$Te$_{4}$. Our research deepens the understanding of the relationship between surface magnetism and the TSS.

%\acknowledgements
This work was funded by the Chinese National Science
Foundation Grant Numbers 12134012. The numerical calculations were performed on the USTC HPC facilities.

%\bibliographystyle{apsrev}
%\bibliography{reference}

\end{document}